# Giant magnetoimpedance in composite wires with insulator layer between non-magnetic core and soft magnetic shell

N.A. Buznikov [a,*], A.S. Antonov [b], A.B. Granovsky [c], C.G. Kim [a], C.O. Kim [a], X.P. Li [d], S.S. Yoon [e]

[a] *Research Center for Advanced Magnetic Materials, Chungnam National University, Daejeon 305-764, Republic of Korea*

[b] *Institute for Theoretical and Applied Electrodynamics, Russian Academy of Sciences, Moscow 125412, Russia*

[c] *Faculty of Physics, M.V. Lomonosov Moscow State University, Moscow 119992, Russia*

[d] *Department of Mechanical Engineering and Division of Bioengineering, National University of Singapore, Singapore 119260, Singapore*

[e] *Department of Physics, Andong National University, Andong 760-749, Republic of Korea*

**Abstract**

A method for calculation of the magnetoimpedance in composite wires having an insulator layer between non-magnetic core and soft magnetic shell is described. It is assumed that the magnetic shell has a helical anisotropy and the driving current flows through the core only. The distribution of eddy currents and expressions for the impedance are found by means of a solution of Maxwell equations taking into account the magnetization dynamics within the shell governed by the Landau–Lifshitz equation. The effect of the insulator layer on the magnetoimpedance is analyzed.

---

[*] Corresponding author. *E-mail address:* n_buznikov@mail.ru

One of the promising materials for applications of the giant magnetoimpedance (GMI) effect is so-called composite wire. These wires consist of a highly conductive non-magnetic inner core and a soft magnetic shell and can be fabricated by means of electrodeposition [1,2], cold-drawn technique [3] and rf-sputtering [4]. Theoretical estimations [5–7] have demonstrated that the presence of the highly conductive core in the composite wires results in significant enhancement of the GMI effect. However, theory predicts higher field sensitivity of the GMI in comparison with that obtained in experiments. Probably, it is related to defects at the core–shell interface reducing magnetic softness of composite wires. One of the ways to improve the magnetic softness may consist in adding of an insulator layer between the core and shell. Studies of multilayered film structures have showed that the GMI effect in films with insulator layers is higher than that in films without insulator layers [8,9]. Recently, it has been demonstrated that the adding of an insulator layer may lead to an increase of the GMI ratio in composite wires [10].

Although the insulator layer prevents the penetration of the driving current into the magnetic shell, the change of the magnetic flux induces eddy currents in the shell. Since the permeability of the shell depends significantly on the external magnetic field, the distribution of the eddy currents in both the core and shell varies with the external field, which results in the field dependence of the impedance. In this work, the effect of the insulator layer on the GMI effect in composite wires is studied theoretically.

Let us consider a composite wire consisting of a highly conductive core of diameter $r_1$, an insulator layer of thickness $t_i$ and a soft magnetic shell of thickness $t_m$. The driving electric field $e=e_0\exp(-i\omega t)$ is non-zero only in the core region, and the external DC magnetic field $H_e$ is parallel to the wire axis. It is assumed that the magnetic shell has a single-domain structure, and the anisotropy axis makes the angle $\psi$ with the transverse direction. The eddy currents distribution and the impedance are found by means of a solution of Maxwell equations taking into account the magnetization dynamics within the shell governed by the Landau–Lifshitz equation. A linear approximation with respect to time variable parameters is used.

The distribution of the electric and magnetic fields within the core, $\rho \leq r_1$, is determined by the driving and eddy currents and can be presented in the form

$$e_z^{(1)}(\rho) = AJ_0(k\rho),$$
$$h_\varphi^{(1)}(\rho) = (4\pi\sigma_1/ck)AJ_1(k\rho),$$
$$e_\varphi^{(1)}(\rho) = BJ_1(k\rho), \qquad (1)$$
$$h_z^{(1)}(\rho) = (4\pi\sigma_1/ck)BJ_0(k\rho),$$

where $\sigma_1$ is the core conductivity, $c$ is the velocity of light, $J_0$ and $J_1$ are the Bessel functions of the first kind, $k=(1+i)/\delta_1$, $\delta_1=c/(2\pi\sigma_1\omega)^{1/2}$, $A$ and $B$ are the constants and the subscripts $\varphi$ and $z$ correspond to the circular and longitudinal components of the fields.

In the region of the insulator layer, $r_1 \leq \rho \leq r_2 = r_1 + t_i$, the fields satisfying the continuously conditions at the core–insulator interface are given by

$$e_z^{(2)}(\rho) = A[J_0(kr_1) + kr_1 J_1(kr_1)\log(r_1/\rho)] - e_0,$$
$$h_\varphi^{(2)}(\rho) = (4\pi\sigma_1/ck)(r_1/\rho)AJ_1(kr_1),$$
$$e_\varphi^{(2)}(\rho) = B[k(\rho^2 - r_1^2)J_0(kr_1)/2 + r_1 J_1(kr_1)]/\rho, \qquad (2)$$
$$h_z^{(2)}(\rho) = (4\pi\sigma_1/ck)BJ_0(kr_1).$$

Within the magnetic shell, $r_2 \leq \rho \leq r_3 = r_2 + t_m$, Maxwell equations can be reduced to two coupled differential equations for the magnetic field components [11,12]. The solutions of these equations have been found for composite wires without the insulator layer for the cases of high and low frequencies [13]. At low frequencies, the solution has a form of series representing the Taylor expansion of the corresponding Bessel functions [13]. In the case of high frequencies, the fields can be expressed as

$$h_\varphi^{(3)}(\rho) = \cos\theta[C\exp(\lambda_1 x) + D\exp(\lambda_2 x)] + \sin\theta[E\exp(\lambda_3 x) + F\exp(\lambda_4 x)],$$
$$h_z^{(3)}(\rho) = \sin\theta[C\exp(\lambda_1 x) + D\exp(\lambda_2 x)] - \cos\theta[E\exp(\lambda_3 x) + F\exp(\lambda_4 x)], \qquad (3)$$
$$e_\varphi^{(3)} = -\frac{c}{4\pi\sigma_3}\frac{\partial h_z^{(3)}}{\partial \rho}, \quad e_z^{(3)} = \frac{c}{4\pi\sigma_3}\frac{\partial h_\varphi^{(3)}}{\partial \rho},$$

where $\lambda_{1,2}=\pm(1-i)/\delta_3$, $\lambda_{3,4}=\pm(1-i)(\mu+1)^{1/2}/\delta_3$, $x=\rho-r_3$, $C$, $D$, $E$ and $F$ are the constants, $\delta_3=c/(2\pi\sigma_3\omega)^{1/2}$, $\sigma_3$ is the conductivity of the magnetic shell, $\theta$ is the equilibrium angle between the magnetization and the transverse direction and $\mu$ is the effective permeability, which are given by [13]

$$H_a \sin(\theta - \psi)\cos(\theta - \psi) - H_e \cos\theta = 0,$$
$$\mu = \gamma 4\pi M /[\omega_1 - i\kappa\omega - \omega^2/(\omega_2 - i\kappa\omega)], \quad (4)$$
$$\omega_1 = \gamma[H_a \cos\{2(\theta - \psi)\} + H_e \sin\theta],$$
$$\omega_2 = \gamma[4\pi M + H_a \cos^2(\theta - \psi) + H_e \sin\theta].$$

Here $H_a$ is the anisotropy field, $M$ is the saturation magnetization, $\gamma$ is the gyromagnetic constant and $\kappa$ is the Gilbert damping parameter.

To calculate the impedance, the fields distribution outside the wire should be found. In the case of the wire excitation by the current, the longitudinal component of the magnetic field in the external region is much less than the circular one [11]. Further, we neglect the radiation term, since it is low at frequencies below several hundreds MHz for the wires of length of the order of 1 cm. Then, the distribution of the fields outside the wire, $\rho \geq r_3$, can be expressed as [6]

$$e_z^{(4)}(\rho) = G \log(l/\rho),$$
$$h_\varphi^{(4)}(\rho) = -i(c/\omega)G/\rho. \quad (5)$$

Here $G$ is the constant and $l$ is the wire length, $l \gg r_3$.

The constants in Eqs. (2), (3) and (5) can be found from the continuity conditions for the fields at $\rho = r_2$ and $\rho = r_3$. The diagonal component of the wire impedance $Z_{zz}$ can be calculated as the ratio of the applied potential difference $le_0$ to the total current flowing through the core [6]:

$$Z_{zz} = le_0 k / 2\pi\sigma_1 r_1 A J_1(kr_1). \quad (6)$$

The off-diagonal impedance $Z_{\varphi z}$ is defined as the ratio of the pick-up coil voltage to the total current in the wire and is proportional to the off-diagonal component of the surface impedance tensor [11,12]:

$$Z_{\varphi z} = (4\pi N/c) \times [e_\varphi^{(3)}(r_3)/h_\varphi^{(3)}(r_3)], \quad (7)$$

where $N$ is the number of turns in the pick-up coil.

Fig. 1 shows the electric field components as a function of the radial coordinate calculated for the wire with the insulator layer. The jump in the radial dependence of the longitudinal field is due to the driving field that is applied to the core only. For comparison, the fields distribution within the wire without the insulator layer is presented in Fig. 1 also.

The longitudinal field within the shell of the wire with the insulator layer decreases from the insulator−shell interface to the wire surface, whereas for the wire without the insulator layer, the field reaches its maximum at the wire surface. On the contrary, the circular component of the electric field has a similar behavior for both the composite wires, since the appearance of the circular electric field is attributed to the cross-magnetization process in the shell and depends slightly on the presence of the insulator layer.

Although the current distribution differs significantly for the wires with and without insulator layer, the diagonal impedance $Z_{zz}$ does not change drastically. At $\psi < \pi/4$, the field dependence of $Z_{zz}$ exhibits symmetrical two-peak behavior. Fig. 2 presents the frequency dependence of the diagonal impedance ratio $\Delta Z_{zz}$ for different values of the insulator layer thickness $t_i$. This ratio is defined as the difference between the peak impedance value and the impedance at zero external field. It follows from Fig. 2 that for the wires with thin insulator layer, the impedance ratio increases slightly in comparison with the wire without the layer. With an increase of $t_i$, the impedance ratio drops due to a decrease of the eddy currents induced in the shell. Note that similar impedance calculations have been made for thin-film magnetic inductors assuming that the permeability is a scalar constant [14,15]. It has been demonstrated that the adding of the insulator layer does not change significantly the impedance up to very high frequencies.

Shown in Fig. 3 is the off-diagonal impedance ratio $\Delta Z_{\varphi z}$ for the wires with and without insulator layer. Note that the off-diagonal impedance is independent of the insulator layer thickness, since the circular electric field in the shell is much higher than that in the core and depends slightly on the insulator layer thickness. The adding of the insulator layer results in a significant growth of the impedance ratio at high frequencies. This fact is attributed to the change in the total current flowing through the wire. At high frequencies, the current drops more sharply in the wire with the insulator layer (see inset in Fig. 3).

It is assumed above that the magnetic properties of the composite wire do not change after the insulator layer adding. In real wires, the adding of the insulator layer may improve magnetic softness of the shell, which leads to an additional increase of the impedance. Probably, this fact may explain the GMI ratio enhancement observed in Ref. [10].

In conclusion, the model to calculate the magnetoimpedance in the composite wires with the insulator layer between the non-magnetic core and magnetic shell is developed. It is shown that the field dependence of the shell permeability effects significantly on the eddy current distribution and leads to the GMI effect. The adding of thin insulator layer results in an enhancement for both the diagonal and off-diagonal impedance at sufficiently high frequencies, which favours such wires for sensor applications.

This work was supported by the Korea Science and Engineering Foundation through ReCAMM and by the Korea Electrical Engineering and Science Research Institute. N.A. Buznikov acknowledges the Brain Pool Program.

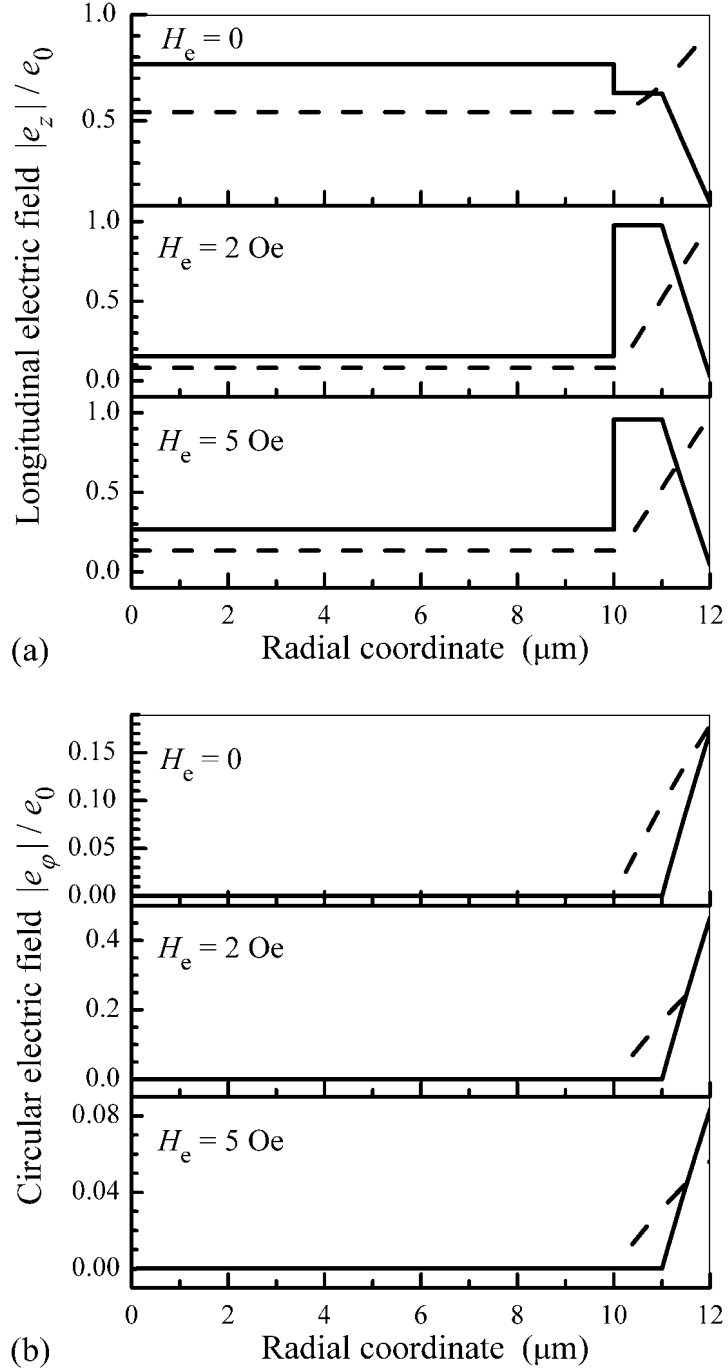

Fig. 1. Dependences of the longitudinal (a) and circular (b) electric field on the radial coordinate at $f=\omega/2\pi=1$ MHz and different external fields. Solid lines, wire with dielectric layer; dashed lines, wire without dielectric layer, $r_2=r_1$. Parameters used for calculations are $r_1=10\,\mu\text{m}$, $r_2=11\,\mu\text{m}$, $r_3=12\,\mu\text{m}$, $l=1\,\text{cm}$, $M=600\,\text{G}$, $H_a=2\,\text{Oe}$, $\psi=0.1\pi$, $\sigma_1=5\times10^{17}\,\text{s}^{-1}$, $\sigma_3=10^{16}\,\text{s}^{-1}$, $\kappa=0.1$.

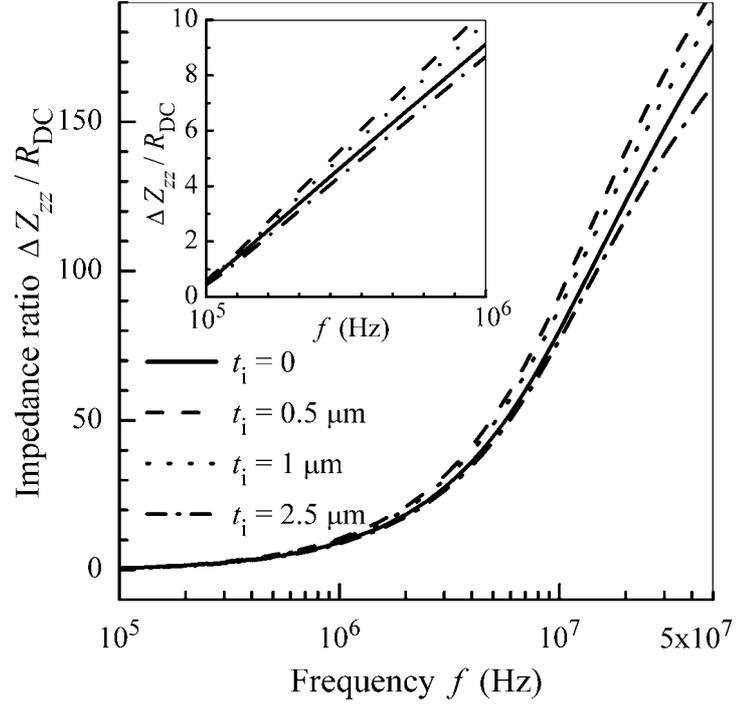

Fig. 2. Frequency dependence of diagonal impedance ratio at different values of the insulator layer thickness, $R_{DC}=l/\pi\sigma_1 r_1^2$. Inset shows the enlarged view for low frequencies. Parameters used for calculations are $r_1=10\,\mu m$, $t_m=2\,\mu m$, $l=1\,cm$, $M=600\,G$, $H_a=2\,Oe$, $\psi=0.1\pi$, $\sigma_1=5\times10^{17}\,s^{-1}$, $\sigma_3=10^{16}\,s^{-1}$, $\kappa=0.1$.

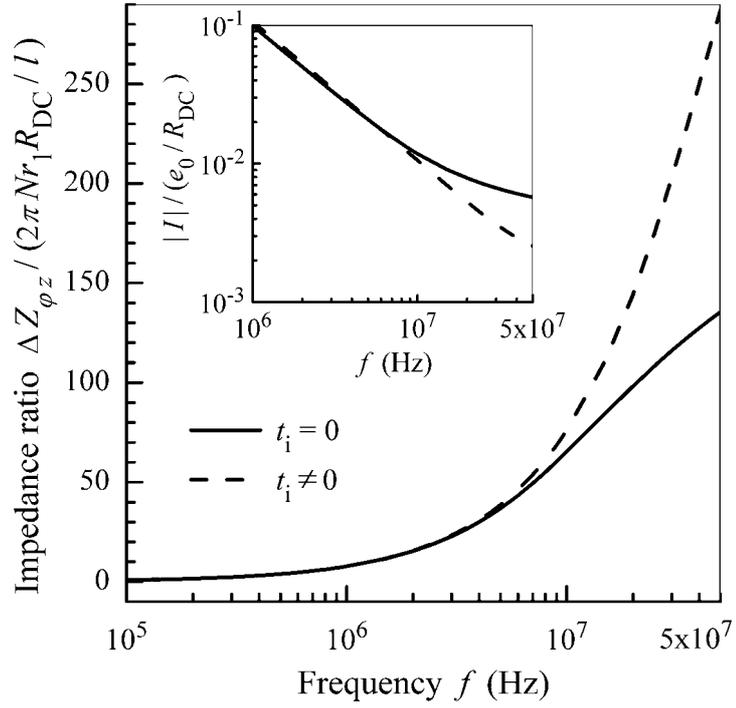

Fig. 3. Frequency dependence of off-diagonal impedance ratio for wires with and without insulator layer. Inset shows total current $I$ versus frequency at $H_e=2$ Oe. Parameters are the same as in Fig. 2.